\documentclass[twocolumn,showpacs,preprintnumbers]{revtex4}
\usepackage{amsmath}
\usepackage{amssymb}
\usepackage{graphicx}
\usepackage{bm}
\begin{document}
\title{Isospin Distillation with Radial Flow: a Test of the Nuclear Symmetry Energy}
\author{M. Colonna$^{a,b}$, V.Baran$^{c}$, M.Di Toro$^{a,b}$,
H.H.Wolter$^{d,a}$}
\affiliation{
        $^a$LNS-INFN, I-95123, Catania, Italy\\
	$^b$Physics and Astronomy Dept. University of Catania, Italy\\
        $^c$NIPNE-HH, Bucharest and Bucharest University, Romania\\
        $^d$Dept. Physik, University of Munich, 85748 Garching, Germany\\}

\begin{abstract}
We discuss mechanisms related to isospin transport 
in central collisions between neutron-rich systems at Fermi energies. 
A fully consistent study of the isospin distillation and expansion dynamics
in two-component systems is presented in the framework of a stochastic transport theory. 
We analyze
correlations between fragment observables, focusing on the study of 
the average N/Z of fragments, as a function of  
their kinetic energy.
We identify an EOS-dependent relation between these observables, allowing
to better characterize the fragmentation path and to access new information 
on the low density behavior of the symmetry energy.  

\end{abstract}
\maketitle

Heavy ion reactions with exotic nuclei at Fermi energies 
can be used to study the
properties of the 
symmetry term at densities below and around the saturation value.
In fact, this regime
of densities is important for studies of the structure of exotic nuclei,
of the neutron star crust, and for supernova explosions, where a key issue is
the clustering of low-density matter \cite{Lat}.  
In central collisions at 30-50 MeV/u, where the full disassembly of the system
into many fragments is observed, one can study specifically properties of 
liquid-gas phase transitions occurring in asymmetric matter in the presence
of radial flow. 
For instance, 
in neutron-rich matter, when two phases co-exist, one expects to observe 
an isospin distillation:
fragments (liquid) appear more symmetric
with respect to the initial matter, while light particles (gas) are 
more neutron-rich \cite{mue95,bao197,rep,rep1}. 
The amplitude of this effect
depends on 
 specific properties of the isovector part of the nuclear interaction, 
namely on the value and the derivative of the symmetry 
energy at low density.  Hence the analysis of the isotopic content
of the reaction products  
allows to get information on the low-density 
isovector Equation of State (EOS) \cite{EPJA}. 

This investigation is interesting in a more general context:
In heavy ion collisions the dilute phase appears during the expansion
of the interacting matter. 
Thus we study effects of the coupling of expansion, fragmentation and distillation in a two-component (neutron-proton) system.
In a statistical model, the effect of the expansion on fragment emission
was studied earlier in Ref.\cite{Fri}.
Here we present a fully consistent dynamical study, based on
microscopic transport approaches widely tested in 
heavy ion collisions \cite{rep,rep1,TWINGO,Salvo,BaranNPA703}.
This leads us to single out the isospin signal as a good tracer of the reaction
mechanism and to suggest new correlation observables 
to probe the symmetry term
at sub-saturation density.

Recently, some efforts have been devoted to the study of the isotopic 
content of
pre-equilibrium emission,
looking  in particular at 
the emitted neutron to proton ratio as a function of the kinetic energy
\cite{Famiano,Fam1}.
In this Letter 
we propose to extend this type of investigation to fragments.

Correlations between fragment charges and velocities
have recently been observed, providing information on the
interplay between thermal and entrance channel (collective) effects in the 
fragmentation mechanism \cite{Indra,Indra1,EPJA_tab}. 
The study of the correlations between fragment isotopic content and
kinematical properties should allow to get
a deeper insight into the reaction path and 
to
study more in detail the effects of different EOS's and, in 
particular, of the symmetry energy on 
fragment properties \cite{Lionti}. 

Theoretically the evolution of complex systems  
can be described
by a transport equation with a fluctuating term, the so-called
Boltzmann-Langevin equation (BLE):
\begin{equation}
{{df}\over{dt}} = {{\partial f}\over{\partial t}} + \{f,H\} = I_{coll}[f] 
+ \delta I[f],
\end{equation}
 where $f({\bf r},{\bf p},t)$ is the one-body distribution function, 
 $H({\bf r},{\bf p},t)$ is the one-body Halmitonian and 
$\delta I[f]$ represents the fluctuating part of the two-body
collision integral \cite{Ayik,Randrup}.
Here we will follow the approximate treatment to the BLE 
presented in Ref.\cite{Salvo}, 
the Stochastic Mean Field (SMF) model, 
that 
consists in the implementation of stochastic spatial density fluctuations.

Calculations have been performed using a Boltzmann-Nordheim-Vlasov (BNV) 
code (TWINGO), 
where the test particle
method is used to solve Eq.(1) \cite{TWINGO}.
We adopt a soft EOS, with compressibility modulus $K = 200 MeV$ and,
for the density  ($\rho$) dependence of the symmetry energy, 
we consider two 
representative parameterizations, 
$E_{sym}(\rho,I)/A \equiv C_{sym}(\rho)I^2, 
I \equiv (N-Z)/A$ : 
one with a rapidly increasing behaviour 
 with density, roughly proportional to $\rho^2$ (asystiff)
and one with a kind of saturation above normal
density (asysoft, $SKM^*$) (see Ref.\cite{rep1,BaranNPA703} 
for more detail).
The two parameterizations obviously cross at normal density. 
The symmetry energy
at densities below the normal value 
is larger in the asysoft case, 
while  above normal density  it is higher in the asystiff case.  
Hence in the low-density regime, that is in the
region of interest for our analysis of spinodal instabilities in central 
collisions, isospin effects are expected to
be stronger in the asysoft case. 


\begin{figure}
\includegraphics[width=8.cm]{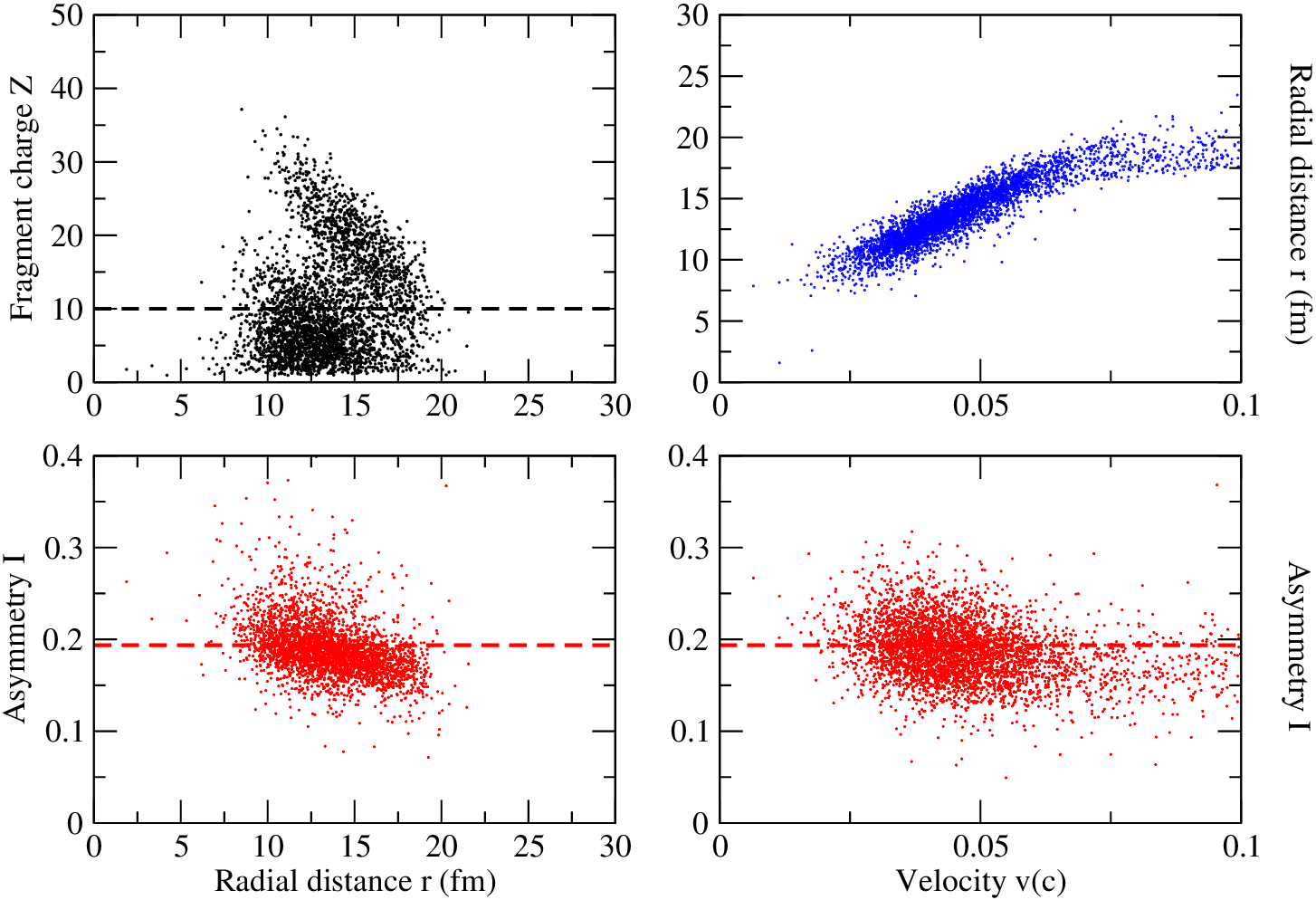}

\caption{(Color online) Top panels: (left) Correlations between fragment charge Z
and radial distance from the system center of mass , for the reaction 
 $^{124}Sn + ^{124}Sn$, at b = 2 fm, E/A = 50 MeV/u (asystiff interaction). 
(right) Correlations between fragment velocity and radial distance. 
Bottom panels: Correlations between fragment asymmetry and 
radial distance (left) or fragment velocity (right). 
The dashed line corresponds to the system initial asymmetry.} 
\label{rapidity}
\end{figure}


We will focus on central collisions, $b = 2~fm$, in symmetric reactions
between systems having three different initial asymmetries: 
$^{112}Sn + ^{112}Sn,^{124}Sn + ^{124}Sn,
^{132}Sn + ^{132}Sn,$ with $(N/Z)_{in}$ = 1.24,1.48,1.64, respectively. 
The considered 
beam energy is 50 MeV/u.
1200 events have been run for each reaction and for each of the two 
symmetry energies adopted. 
The first two reactions, $^{112}Sn + ^{112}Sn$ and $^{124}Sn + ^{124}Sn$
have been widely  investigated both from the experimental 
and theoretical point of view \cite{tsangprl1,BaranNPA703,bao,liu}. 

We first focus on some general properties of the fragmentation events, 
as described by the SMF model.
Along the fragmentation path
a bubble-like configuration is formed, where the initial 
fragments are located  \cite{rep}. 
The average fragment multiplicity is approximately equal to 6 
for the reactions considered here \cite{BaranNPA703}.
Along the reaction path, several nucleons are emitted at the early stage
(pre-equilibrium emission) and/or are evaporated while fragments are formed. 
Primary fragments are identified by applying 
a coalescence procedure to the matter with density larger than 
$\rho_{cut} = 1/5~\rho_0$ (liquid phase).
The remaining particles are considered as belonging to the gas phase.
The correlation between fragment charge  and c.m. position, 
as calculated at the ``freeze-out'' time,  
is displayed in Fig.1 (top left panel) 
for the  $^{124}Sn + ^{124}Sn$ system (asystiff interaction).
The ``freeze-out'' time is defined as the time when fragments are completely
formed and their multiplicity does not evolve anymore.   
One observes that fragments are located on a shell 
of about 10-15 fm radius and consist of 
two components. 
The upper component can be associated with events
where one or two larger fragmemts survive, as a memory of the entrance channel.
The fragment charge shows a decreasing trend 
with the distance.
The second component can be related to more explosive events, where
the system breaks up into a larger number of smaller fragments. 
In simulations of very central collisions 
(b = 0 fm) only this second component is present.
Since we are mostly interested in this component, 
fragments with charge larger than 10 (dashed line in the figure) will not
be considered in the following analysis.  

Another feature observed at this beam energy is the existence of
a radial collective flow \cite{Indra}, 
as evidenced in Fig.1 (top right panel)
where the correlation between fragment velocity and radial distance
is shown.   
As a consequence of the fact that large fragments are preferentially
located at a closer distance, one expects to see a decrease of 
the average fragment velocity with the fragment charge. 
This feature has been experimentally observed in central collisions 
\cite{EPJA_tab}. 

The aim of this Letter  is to investigate correlations between fragment 
asymmetry and velocity (or kinetic energy).
The idea 
is that fragmentation originates from the break-up of a composite source
that expands with a given velocity field.  
Since neutrons and protons experience different forces, 
one may expect a different radial flow for the two species. 
In this case,  the N/Z composition of the source would not be uniform, 
but would depend on the radial distance from the center or mass or, 
equivalently, on the local velocity.
This trend should then be reflected in the fragment 
asymmetries as a function of their kinetic energy.  
The existence of such correlations can be qualitatively seen in Figure 1
(bottom panels), where the fragment asymmetry, $I$, is plotted
as a function of the radial distance and velocity.   
It is observed that 
the asymmetry decreases
with the radial distance, indicating a different proton/neutron
radial distribution  in the fragmenting source (bottom left). 
Since radial distance and velocity are correlated (top right), 
this implies 
correlations between fragment asymmetry and velocity,
as seen in the bottom right part of Fig.1.
However, in both cases,  
fluctuations are rather large.

Let us now 
discuss  
the isotopic content of fragments and
emitted nucleons, as obtained with the two iso-EOS considered.
In the following we will restrict our analysis to fragments with charge
between 3 and 10 (intermediate mass fragments 
(IMF) ).
The average N/Z of emitted nucleons (gas phase) and IMF's
is presented in Fig.2 as a function of the initial $(N/Z)_{in}$ 
of the three colliding systems.
\begin{figure}
\includegraphics[width=8.cm]{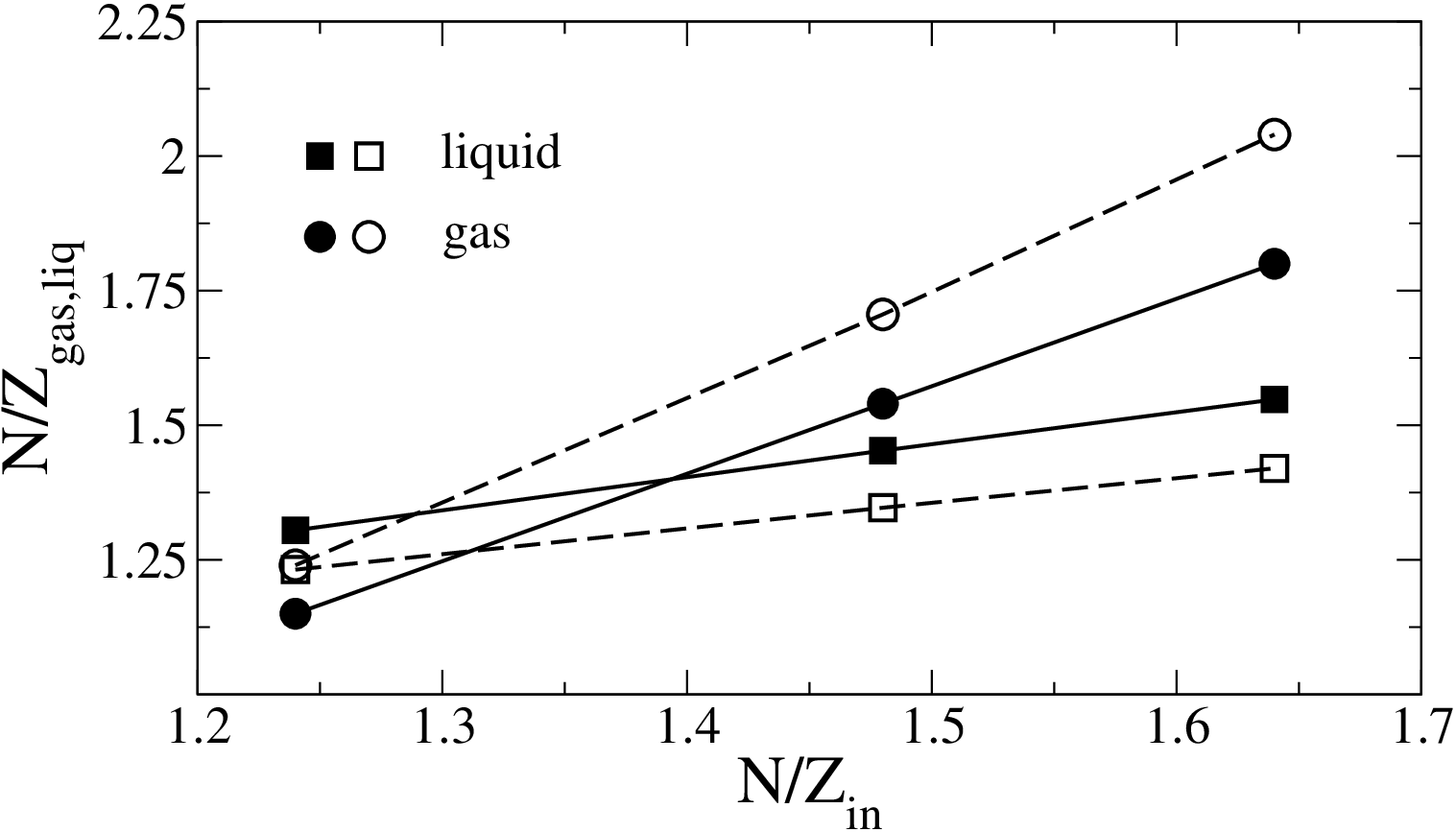}
\caption{The N/Z of the gas (circles) and of the liquid (squares) phase
is displayed as a function of the system initial N/Z.
Full lines and full symbols refer to the asystiff parameterization. Dashed
lines and open symbols are for asysoft. 
} 
\end{figure} 
Generally   
the gas phase is seen to be more neutron-rich in the
asysoft relative to the asystiff 
case, while IMF's are more symmetric. 
The difference between the asymmetries of the gas and liquid phases 
increases
with the $(N/Z)_{in}$ of the system, and is always larger in the 
asysoft case, 
due to the larger value of the symmetry energy at low density
\cite{BaranNPA703}. 
In the asystiff case, however, due to the rather 
low value of the symmetry energy,
this difference is negative for the neutron-poor
system ($^{112}Sn + ^{112}Sn$). 
In fact, in this case Coulomb effects dominate 
and protons are
preferentially emitted. 

Now we move to discuss the 
correlations between fragment isotopic 
content and kinematical properties. 
As a measure of the isotopic composition of the IMF's we will consider
the sum of neutrons, $N = \sum_i N_i$, and protons, $Z = \sum_i Z_i$, 
of all IMF's in a given kinetic energy bin (here taken as 
$1.5~MeV/u$), 
in each event. Then we take the ratio $N/Z$ and we consider
the average over the ensemble of events. 
This observable is plotted in Fig.3 for the three reactions 
and the two iso-EOS considered and is seen to be rather sensitive.  
For the neutron-poor system, the N/Z decreases with the
fragment kinetic energy, expecially in the asystiff case. Here the
symmetry energy is relatively low at low density \cite{BaranNPA703} and
the Coulomb repulsion
pushes the protons towards the surface of the system. Hence, more
symmetric fragments acquire larger velocity.
The same effects are responsible for the relatively 
neutron-poor pre-equilibrium emission in this case (see Fig.2).  
The decreasing trend is less pronounced in the asysoft case
because now Coulomb effects are counterbalanced by the larger
attraction of the proton symmetry potential. 
In systems with larger initial asymmetry, the decreasing
trend is inversed, due to the larger neutron repulsion in neutron-rich
systems.
Due to the balance between the asymmetry of the liquid phase, that is 
larger in the asystiff case, and the value of the symmetry energy at 
low density, that is larger in the asysoft case, the last effect is
of similar amplitude in the two EOS's.

In order to isolate isospin effects, it is advantageous to compare results
for systems with different initial asymmetry \cite{tsangprl1,Fam1}. 
Hence we construct the ``double'' $N/Z$ ratio, as a function of the
fragment kinetic energy, by taking the ratio of the $N/Z$ 
of a neutron-rich relative to a neutron-poor reaction. 

The double $N/Z$ ratio was already investigated for the
pre-equilibrium emission, where experimental data have been compared to
the predictions of the BUU transport model \cite{Famiano}. 
A larger value of the double ratio was observed in the asysoft case,
corresponding to the larger amount of emitted neutrons in neutron-rich
systems, 
expecially at
high kinetic energy.
This is qualitatively in agreement with our findings, as seen 
in Fig.2.
\begin{figure}
\includegraphics[width=8.cm]{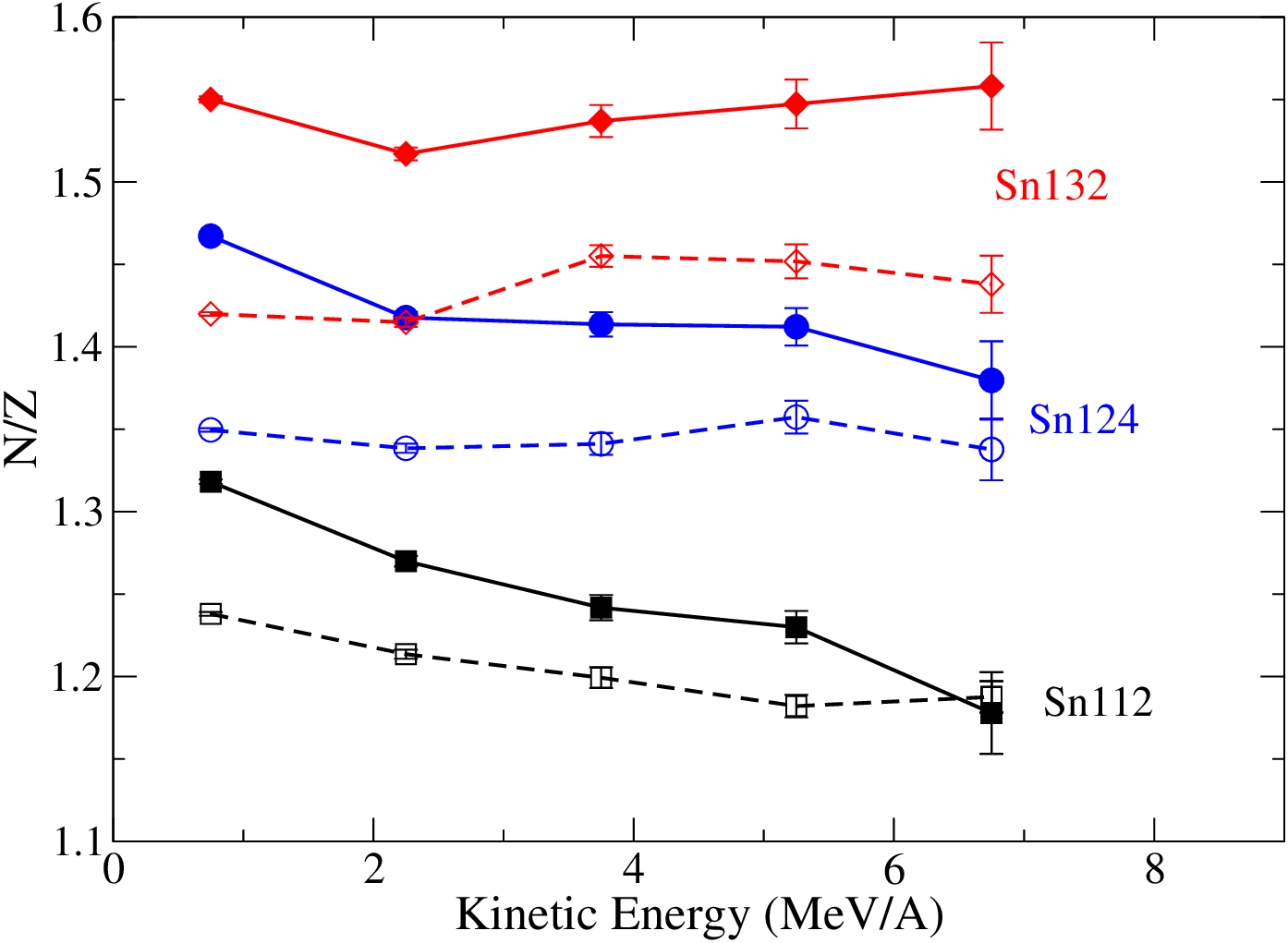}
\caption{
(Color online) The fragment $N/Z$ (see text) as a function of the kinetic energy. 
Lines are as in Fig.2 and the symbols distinguish the different initial 
systems.
} 
\end{figure}
In Fig.4 (top panel) we show the double $N/Z$ ratio constructed for
the pair of reactions  $^{124}Sn + ^{124}Sn$ and $^{112}Sn + ^{112}Sn$ (circles)
and the pair  $^{132}Sn + ^{132}Sn$ and $^{112}Sn + ^{112}Sn$ 
(squares), respectively. 
Conversely to what was observed for the pre-equilibrium emission 
\cite{Famiano}, the fragment
double $N/Z$ ratio is larger in the asystiff case. 
This reflects the fact that
fragments originate from the composite system that survives
the pre-equilibrium emission, the asymmetry of which 
is larger in the asystiff case, expecially in neutron-rich systems.   
>From this point of view, 
fragments bear a complementary information with respect to the pre-equilibrium
emission. Hence
opposite effects are expected and observed in Fig.4 
for fragment emission.  
The differences between the two iso-EOS's increase with the very neutron-rich
reaction $^{132}Sn + ^{132}Sn$,  
as expected,
however the sensitivity to the iso-EOS appears rather
small (less than 5$\%$ ). 

        To  construct a more sensitive observable, one can combine 
the two main isospin effects represented in Fig.3:
1) the decreasing trend observed for the lighter system, with a 
rather EOS-dependent slope;
2) the change of slope, up to the inversion of the decreasing trend,
in the neutron-richer systems. 
Thus we propose to consider a ``shifted'' ratio, $(N/Z)_s$, 
by taking the asymmetry relative to the lowest kinetic energy bin, which 
is related to the slope of the curves 
in Fig.3, which is dependent on the 
iso-EOS's considered. Thus we expect to enlarge the sensitivity of the ratio
to the symmetry energy.    
\begin{figure}
\includegraphics[width=8.5cm]{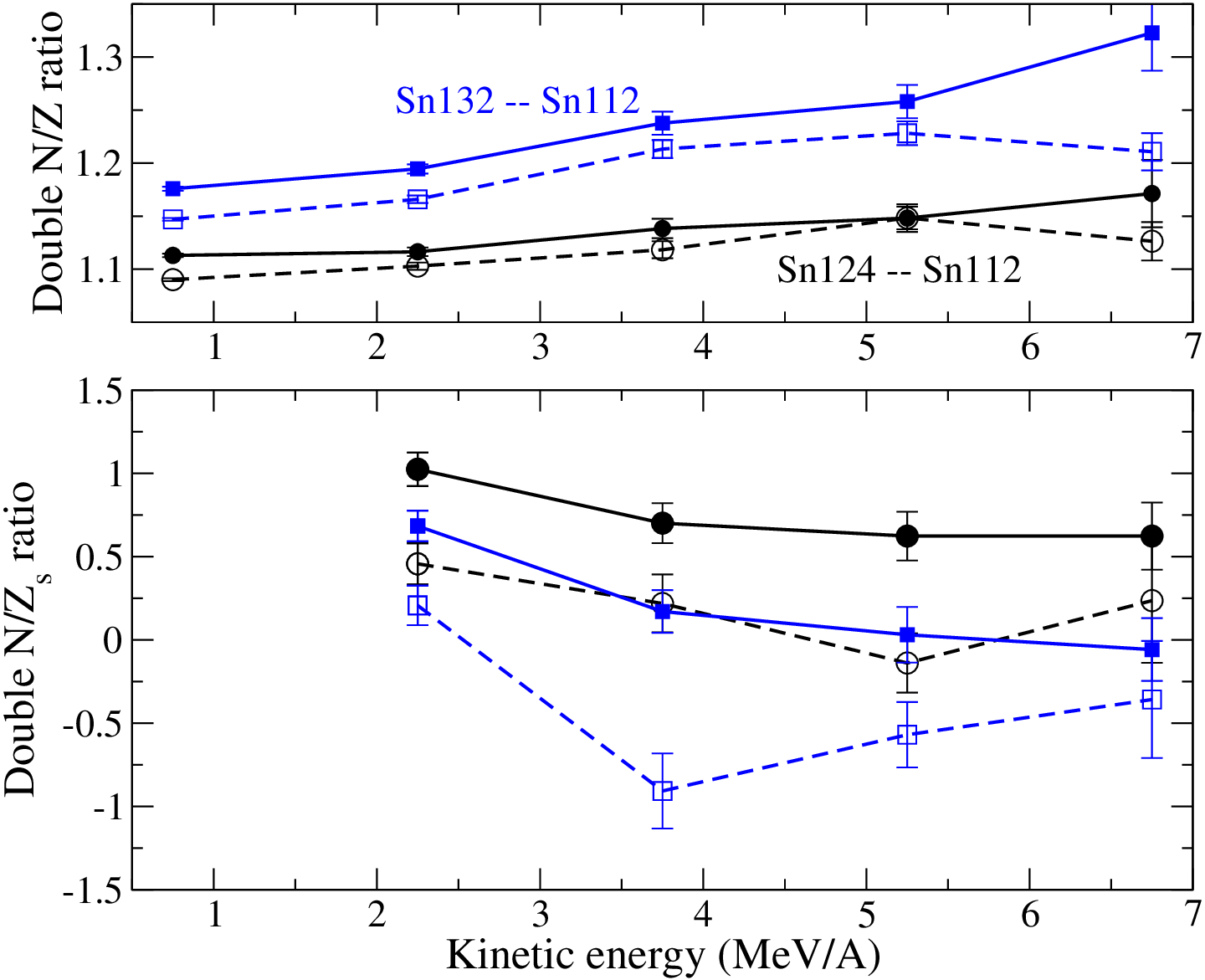}
\caption{(Color online) Top: The double $N/Z$ fragment ratio as a function of the kinetic energy
for the reactions $^{124}Sn + ^{124}Sn$ to $^{112}Sn + ^{112}Sn$
(squares) and the reactions $^{132}Sn + ^{132}Sn$ to $^{112}Sn + ^{112}Sn$ 
(circles). Full lines and symbols are for the asystiff, dashed lines and
open symbols for the asysoft EOS.
Bottom: Same as in the top panel, but for the shifted N/Z ratio
(see text).   } 
\end{figure}
In Fig.4, bottom panel, we present the behaviour of the double $(N/Z)_s$ ratio
for the pairs 
of reactions  considered before, 
and for the two iso-EOS's.
The double ratio constructed in this way is rather sensitive to
the EOS employed, even for the pair of reactions 
with the closest $(N/Z)_{in}$ ($^{124}Sn + ^{124}Sn$, $^{112}Sn + ^{122}Sn$).
Lower values are obtained in the asysoft case. 

We can try to interpret the results in Fig.4 in a simple model, by approximating the ratios in Fig.3 by a linear relation: $(N/Z)_i(E_{kin}) = (N/Z)_i(0) + 
m_i E_{kin}$, where $i$ refers to the system. From Fig.3 we observe that 
$(N/Z)_i(0)$ is roughly shifted by a constant amount for soft and stiff
iso-EOS's. On the other hand, the slopes in Fig.3 for the soft, relative
to the stiff iso-EOS, corresponds roughly to a rotation, such that the
difference $\Delta m_{2,1}$ between two systems is almost independent of the
iso-EOS. The shifted double $N/Z$ ratio between two systems is then the ratio of the 
corresponding slopes and constant in energy. This is also roughly confirmed
in Fig.4. Furthermore it can be written as: 
 $DR_{2,1} = 1 +  \Delta m_{2,1} /m_1$.
>From this relation it is clear that the deviation of the shifted double ratio
from $1$ increases with the difference $\Delta m_{2,1}$ and with the flattening
of the slope $m_1$ (notice that in our simulations $m_1$ is always negative).  
The dependence on the iso-EOS then lies in the slope of the neutron-poorest
system, which is most strongly affected by the symmetry energy. 
So, the shifted $N/Z$ ratio allows to access directly  
the sensitivity of the slope $m_1$ to the EOS. 
We also remark that the usual double ratio, displayed in the top panel in
Fig.4, does not show a large sensitivity to the iso-EOS, because it is dominated by the differences in $(N/Z)_i(0)$, which are relatively small (about $10\%$). 


Here we have considered primary fragments. However, the fragment properties
measured
in experiments are affected by secondary decay. 
While 
the curves presented in Fig.3 could be affected by 
secondary decay, 
we expect, however, the shifted double $(N/Z)$ ratio  
to be more robust.
With the assumption that the average
$N/Z$ of final fragments, $(N/Z)_{fin}$, is reduced by a given constant quantity
$b$ or by a constant factor $a$ with respect to the $N/Z$ of primary
fragments,  
i.e. $(N/Z)_{fin} = a(N/Z - b)$,
which is reasonable for fragments from similar reactions at the same energy
(i.e. similar excitation energy),  
the shifted double $(N/Z)$ ratio would not be 
affected at all by secondary decay effects.  

In conclusion, we have presented a 
new analysis of fragmentation reactions
that would allow to extract relevant information on the low-density behaviour
of the symmetry energy.  We 
propose an
EOS-dependent relation between the $N/Z$ and the kinetic energy of
IMF's,  which 
is linked to the different forces experienced
by neutrons and protons along the fragmentation path, which in turn depend
on the details of the isovector part of the nuclear interaction. 
The isospin distillation mechanism in fragment formation 
appears naturally coupled to
the underlying expansion dynamics. 
As a consequence, the N/Z composition of the fragmenting source is not
uniform. 
This analysis can be considered as complementary to the 
pre-equilibrium emission studies \cite{Famiano}. 
A parallel investigation of pre-equilibrium and fragment emissions 
would be very important for 
a cross-check of model predictions 
against experimental observables sensitive to different phases 
of the reaction. 

\end{document}